\documentclass[aps,prl,twocolumn]{revtex4}
\usepackage{amsmath,bm,epsfig}

\def\Fbox#1{\vskip1ex\hbox to 8.5cm{\hfil\fboxsep0.3cm\fbox{%
  \parbox{8.0cm}{#1}}\hfil}\vskip1ex\noindent}  

\newcommand{\eq}[1]{(\ref{#1})}
\newcommand{\Eq}[1]{Eq.~(\ref{#1})}
\newcommand{\Eqs}[1]{Eqs.~(\ref{#1})}
\newcommand{\Fig}[1]{Fig.~\ref{#1}}

\newcommand{\B}[1]{{\bm{#1}}}
\newcommand{\C}[1]{{\mathcal{#1}}}    
\let \= \equiv \let\*\cdot \let\~\widetilde \let\^\widehat \let\-\overline

\renewcommand{\sb}[1]{_{\text {#1}}}  
\renewcommand{\sp}[1]{^{\text {#1}}}  

\def\<{\left\langle}    \def\>{\right\rangle}
\def\({\left(}          \def\){\right)}
 \def \[ {\left [} \def \] {\right ]}

\begin{document}
\title{Random Vortex-Street Model for a Self-Similar Plane Turbulent Jet}
\author{ Victor S.
L'vov, Anna Pomyalov, Itamar Procaccia}
\affiliation{Department of Chemical Physics, The Weizmann
Institute of Science, Rehovot 76100, Israel, \ \rm and
}
\author{Rama Govindarajan }
\affiliation{Engineering Mechanics Unit JNCASR, Jakkur
Bangalore 560064, India}
\date{\today}
\begin{abstract}
We ask what determines the (small) angle of turbulent jets. To answer
this question we first construct a deterministic vortex-street model
representing the large scale structure in a self-similar plane
turbulent jet. Without adjustable parameters the model reproduces the
mean velocity profiles and the transverse positions of the large scale
structures, including their mean sweeping velocities, in a
quantitative agreement with experiments.  Nevertheless the exact self
similar arrangement of the vortices (or any other deterministic model)
necessarily leads to a collapse of the jet angle. The observed (small)
angle results from a competition between vortex sweeping tending to
strongly collapse the jet and randomness in the vortex structure, with
the latter resulting in a weak spreading of the jet.

\end{abstract}

 \maketitle

\noindent{\bf Introduction:} We address the apparent universality of
the self-similar structure of plain turbulent free  jet. It had been amply documented\cite{65brad,76GW,82OG,Pope,00GT,86TB,81GYO} that
such jets contain dynamically dominant large-scale self-organized motions
consisting of  two lines of large-scale
vortices centered at staggered position on the two sides of the jet:
$
y(x)= \pm a \, b(x)\,,\quad b(x) = \alpha\,  x\,, \  a\approx 0.85$,
where  $x$ is
 a stream-wise coordinate,  $b(x)$ is the jet half-width~\cite{footnote} and $\alpha$ is the
jet angle. This structure carries about 75\% of the total kinetic
energy \cite{00GT}. Importantly, the remaining turbulent energy
randomizes in part the positions and amplitudes of the coherent
vortices.

In this Letter we first construct   a deterministic vortex-street model
that  excellently reproduces the  experimentally observed
longitudinal mean velocity profile  without any adjustable parameter. The model  predicts  the transverse positions of the large scale structures, including their mean sweeping velocities, in a
quantitative agreement   with experiments. Nevertheless both the existence of a
regular array of vortices and the turbulent randomization are crucial
for understanding  the (relatively small) observed angle $\alpha\sb{exp}\approx 0.1$ of the jet
spreading. Therefore, secondly we  develop a  random-vortex street model, that leads  to an estimate of the jet angle   $\alpha\sb{mod}\approx 0.13 $ (reasonably close to the observed angle)  as a result of  a competition between the regular  and the stochastic components.\\

\noindent{\bf The deterministic model:}  To model the coherent structure of a
plane jet we employ vortices with a finite core size
\cite{82OG}.  Here we choose a simple algebraic model for the vorticity $\omega(r)$ and angular
velocity $v_\varphi(r)$ that allows an analytic treatment:
 \begin{equation}\label{alg}
 \omega(r)=\frac{\Gamma c^4}{\pi (r^4+c^4)^{3/2}}\ \Rightarrow \ v_\varphi(r)= \frac{\Gamma \,r}{2\pi\sqrt {r^4+ c^4}}\ ,
 \end{equation}
where $r$ is the distance from the vortex center. To characterize the self-similar structure  of the jet we enumerate the  vortices by their sequential number $ -\infty  <  n < \infty$ and introduce a  scaling parameter $\lambda>1$ such that the $x$-position of vortex centers are
$ x_n=x_0 \lambda^n$, where $x_0$ is the stream-wise position of a reference vortex.
The $y$-position of the vortices is chosen at the points of inflection of the
mean velocity profile where the production of vorticity has its maximum:
$y_n=(-1)^n a\,\alpha \,   x_n$, where both $a$ and $\alpha$ need to be determined below.

Note that we construct the jet exactly self-similar; in experiments this is
achieved only asymptotically away from the orifice, about 30 times the
width of the orifice.  To flush out the scaling properties of the mean
stream-wise velocity $V_x(x,y)$ and the velocity circulation
$\Gamma_n$ notice that for the free jet the stream-wise momentum flux
$M$ is $x$-independent:  
\begin{equation} \label{M}
M\=\int_{-\infty}^\infty V_x(x,y)^2 dy \approx~ V_x(x,0)^2 b(x)\ .
 \end{equation}
Since $b(x)$ scales as $x$, for constant $M$ the centerline velocity
$V_x(x,0)$ should necessarily scale like $1/ \sqrt{x}$.  Similarly,
one estimates $\Gamma_n\sim V_x(x_n,0) b(x_n)\propto \sqrt{x_n}\propto
\sqrt{\lambda^n}$. Therefore
$ \Gamma_n=\gamma (- \sqrt{\lambda})^n$,
with some dimensional constant $\gamma$.

Note that this model, if used as initial condition for the Euler equation, will lose its self-similarity immediately. Dynamically suitable model that reflects the self similarity of the observed jet will call for a very very much more detailed and parameterized description. The analytic simplicity of the present model results from the fact that the self-similarity should be interpreted statistically rather than dynamically;  in the real jet situation we must allow for the
coalescence or sometime creation or death of vortices.

At an arbitrary position $x,\ y$ and time $t$ the stream-wise velocity field, produced by all the vortices, is:
\begin{subequations}\label{U}
\begin{eqnarray}\label{Ua}
U_x(x,y)&=&\frac{\gamma}{2 \pi} \sum_n   { -(-\sqrt\lambda)^n Y_n }\big / { R_n^2 }\,,\\
R_n ^2&\=& \sqrt{\({X_n}^2+ {Y_n}^2\)^2+c^4}\,, \\
X_n &\=&x-x_n\,, \quad 
 Y_n \= y-y_n\,, 
 \end{eqnarray}\end{subequations}
where    the core radius of the $n$-th vortex $c_n= c_0
\lambda ^n$. Similarly one gets an equation for the
transverse velocity $U_y(x,y)$ by replacing $-Y_n\to X_n$ in the
numerator of \Eq{U}.

We define the {\em mean} (in time) velocity profiles,
$V_x(x,y)\equiv \langle U_x(x,y)\rangle $ and $V_y(x,y)\equiv \langle
U_y(x,y)\rangle$ and the normalized coordinates $\xi\= x/ b(x),
\zeta\= y/ b(x)$.  As the next
step we define a normalized self-similar mean velocity, which is $x$-independent
in the self-similar regime:%
   \begin{equation}\label{Vddag}
   \B V^\ddag ( \zeta)\equiv \B V(x, y/b(x)) /V_x(x,0)\, .
   \end{equation}

\begin{figure}
\centering
\includegraphics[width=0.4\textwidth]{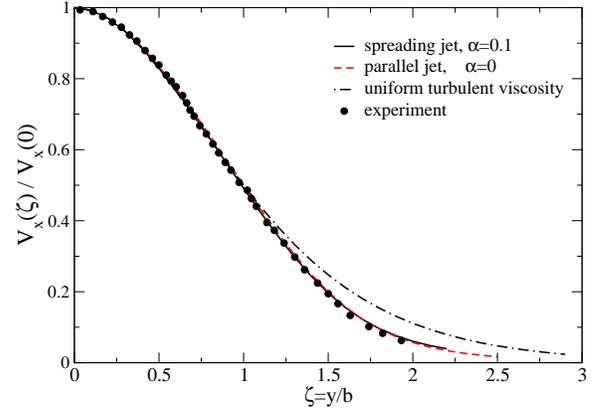}\vskip -.53cm
\caption{Color online. Profiles of the normalized stream-wise mean
velocity: Black squares -- experimental points~\cite{00GT}, black
dashed line -- profile~\eq{Pope-prof} obtained from the Prandtl
closure with constant turbulent viscosity approximation, red
dash-dotted line line -- \Eq{PVS} ($\alpha\to 0$ limit) with
$a=0.747$, $c\approx 1.1$.  Black solid line -- profile~\eq{Ux-fin}
with $\alpha=0.1$ and the same values of $a$ and $c$.}
\label{f:profiles}
\end{figure}

Since the actual experimental value of the spreading angle $\alpha\sb{exp}$ is
small, and since we will find below that the profile of the
longitudinal mean velocity is very weakly dependent on $\alpha$, it is
very worthwhile to proceed and analyze precisely the limit $\alpha\to
0$, where the vortices are arranged on two parallel lines $y=\pm
a$, separated in the streamvise direction by $d$. There we can simplify dramatically the calculation of the mean
stream-wise velocity from Eq. (\ref{U}), in the form of a single
integral. For $\alpha \to 0$ we write $\lambda^n\to 1+(\lambda-1)n$
and substitute in Eq. (\ref{U}) $x_n=x_0+n \,d \, b $,
$d=(\lambda-1)x_0$, $y_n=(-1)^n a b $, $c_n=c \, b$, where $b$ is the
$x$-independent jet half-width:
\begin{subequations}\label{actU}
\begin{eqnarray}
\C U_x(x,y)  =  \frac{\gamma}{2\pi b} \sum_{n=-\infty}^\infty \frac{\C Y_n}{\C R_{n} }\,,&&  \C Y_n\=a-(-1)^n\zeta\,,
\label{actUa}\\ 
\C  U_y(x,y)  = \frac{\gamma}{2\pi b }\sum_{n=-\infty}^\infty  \frac{\C X_n}{\C R_{n} } ,&& \!\!\! \C X_n\= (-1)^n (\xi- nd) 
 \,,    ~~~~\label{actUb} \\
 \C R^2_{n}& \= &  \sqrt{ \(\C X_n^2+\C Y_n^2\)^2+c^4}\ .
\end{eqnarray}
\end{subequations}
The average with respect to time was
replaced by averaging in the $\xi$ direction over one period of
oscillations (between two consecutive vortices at $-d $ and $d $). In
each term in the sum (separately for odd and even $n$) we can change
the integration variables $\xi' =\xi +2\, d\, n$ and eventually
collapse the whole sum into two integrals, finding for $\C V_x (b \,\zeta  ) \equiv \langle \C U_x(x,y)\rangle$:
\begin{subequations} \begin{eqnarray}\label{meanVelU}
\C V_x (b \,\zeta  ) &=&  \frac{\gamma}{4\pi b\,  d}[ I^+(\zeta)+I^-(\zeta)]\,,  \\
I^\pm(\zeta )&=& \int  _{-\infty}^{\infty}A_\pm d\xi\,, \\
A_\pm &\equiv& {(\zeta \pm a ) }\big /
{\sqrt{[(\zeta \pm a )^2+ \xi^2 ]^2+ c^4}}\ .
\end{eqnarray}\end{subequations}%
Notice that  $\C V_y (\zeta)$
vanishes in the limit $\alpha\to 0$ due to symmetry. While these results were
developed for the limit $\alpha \to 0$, we will demonstrate that they
pertain excellently well also for small $\alpha$.
The normalized profile according to \Eqs{Vddag} and \eq{meanVelU}:
\begin{equation}\label{PVS}
\C V_x^\ddag(\zeta)\={[ I^+(\zeta)+I^-(\zeta)]}/ {[ I^+(0)+I^-(0)]} \ .
\end{equation}
 This profile depends only on $a$ and $c$.  With the definition of the
 width of the jet, $b(x)$, we now demand $\C V_x^\ddag(1)=\frac 12$, giving us
 one relation between $a$ and $c$. The second relation between these parameters is
 determined by the position of the inflection point: $d^2 \C V_x^\ddag(\zeta)/d\zeta^2=0$ at
 $\zeta=a$. Solving the two conditions together we find $a=0.747$
 and $c=1.1$. The resulting profile (\ref{PVS}) with these values of the parameters is shown
 in Fig. \ref{f:profiles}.
This profile $\C V_x^\ddag(\zeta)$ obtained for $\alpha=0$ should be compared to the profile
$ V_x^\ddag(\zeta)$
computed for finite $\alpha$ \cite{note1}:
\begin{eqnarray}\label{Ux-fin}
 V_x^\ddag(\zeta)&=&
{[ J^+(\zeta)+J^-(\zeta)]}\big /  {[ J^+(0)+J^-(0)]}\,, \\ \nonumber
J^\pm(\zeta)&\=&\int\limits _0^\infty \frac{d\chi }{\alpha \sqrt \chi }\, \frac{a\chi \pm \zeta}{\sqrt {[(a\chi  \pm  \zeta)^2+(\chi -1)^2/\alpha^2]^2+ c^4 \chi ^4}}\ .
 \end{eqnarray}
 In \Fig{f:profiles} we compare these two profiles and find that as expected they are practically
 indistinguishable for the experimental value $\alpha_{\rm exp} =0.1$. Moreover, the profile \Eq{Ux-fin} is almost insensitive to $\alpha$ for  small $\alpha\lesssim 0.15$.  In the same figure
 we compare the model profile to the experimental data~\cite{00GT}, and find excellent agreement.
 In contrast, the profile
\begin{equation}
\label{Pope-prof}
 V_{{\rm utv},x}^\ddag(\zeta) =\cosh^{-2} \Big[ \zeta \,\ln(1+\sqrt{2})\Big ] \ ,
\end{equation}
computed using  the  ``uniform turbulent viscosity" assumption~\cite{note2}, deviates
strongly from the data for $\zeta>1$.

  Moreover, the present model fits the data in more than one
way. One can actually measure the transverse positions $y_n$ of the
large scale vortices in units of the $b(x_n)$. In the self-similar
regime of the jet the ratio $y_n/b(x_n)$ is $n$-independent and best
estimates~\cite{00GT} give approximately 0.85 in a reasonable agreement with our
value of $a\approx 0.75$. Finally, the measured value of the stream-wise velocity
of the coherent structures in units of the centerline velocity is
about $0.60-0.65$ \cite{86TB,81GYO}. Our model predicts $ \C V_x^\ddag( a)
\approx 0.57$. We remind the reader that all this agreement is
achieved without using any experimental data.

\begin{figure}
~\hskip -1.0 cm
\includegraphics[width=0.55\textwidth]{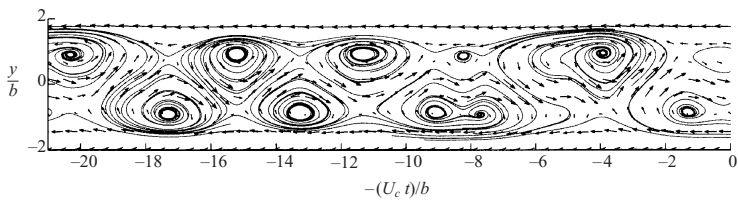}
\caption{Vortex structure (a fragment from experimental proper
orthogonal decomposition, Ref.~\cite{00GT}).}
\label{structure2}
\end{figure}

Having modeled so successfully the stream-wise velocity, we can now
evaluate the transverse velocity by using the incompressibility
constraint and \Eq{Vddag}.  We derive
\begin{equation}\label{Vy}
  V_y^\ddag (\zeta)=\frac{\alpha} 2  \int _0 ^\zeta d\~ \zeta\,
  \Big [  V_x^\ddag (\~ \zeta) + 2\, \~ \zeta \,   {d \,   V_x^\ddag (\~ \zeta)}\big / {d \~ \zeta }\Big]\ .
\end{equation}
 For small $\alpha$ one can neglect the $\alpha$ dependence of the
integrand in the right hand side of \Eq{Vy}, replacing $V_x^\ddag$ by its $\alpha
\to 0$ limit $\C V_x^\ddag$. Now we can compute the ratio
$\alpha'\equiv V_y(x,y)/V_x(x,y)$ at $y=
 a \, b(x)$, finding that the sweeping angle of the vortices
 $\alpha'\approx 0.3\, \alpha$ is considerably smaller than the angle
 $a\, \alpha\approx 0.75 \, \alpha$ that is required to preserve the angle self-similarly.
This result means that the {\em mean} velocity of
 the vortices is not maintaining the self-similar structure at {\em
 any (small) value of $\alpha$}, tending to collapse the structure to
 two parallel lines. We expect this result to be generic for any deterministic model.

\noindent{\bf Random Vortex-Street Model}. Now we have to account for
the fact that the real jet does not display ideal vortex-street
structure, see, e.g.  \Fig{structure2}. The parameters of individual
vortices fluctuate, and we also see the coalescence of vortices as
mentioned above. Moreover, pairs of vortices have finite life-time, as
we see in \Fig{structure2}, that two vortices are about to disappear
at $U_c t\approx -8 \, b$. We reiterate that the finiteness of the
vortex life-time is \emph{required by the self-similarity}, i.e. 
 the sweeping velocity  is 
oriented along the line that connects the vortex centers. Indeed,
the sweeping velocity scales like $ 1/ \sqrt x$, the vortex density
scales like $1/x$, therefore the flux of the vortex number scales like
$ 1/x^{3/2} $ i.e. decreases with $x$.  This means that vortices
disappear during their sweeping downstream, i.e. they have finite life
time, due to their coalescence, or decay, or whatever. It means, as
said above, that self-similar jet structure must be understood in the
statistical sense and randomness of the vortex parameters is a
necessarily element of the jet structure. The relative success of the
deterministic vortex-street model was possible because fluctuations of
the vortex parameters have little consequences on the mean values;
they can be neglected in zeroth-order approximation in describing the
mean velocity profiles. This is not the case for the spreading angle
$\alpha$, which tends to collapse in the deterministic limit.

\begin{figure}
\centering
\includegraphics[width=0.4\textwidth]{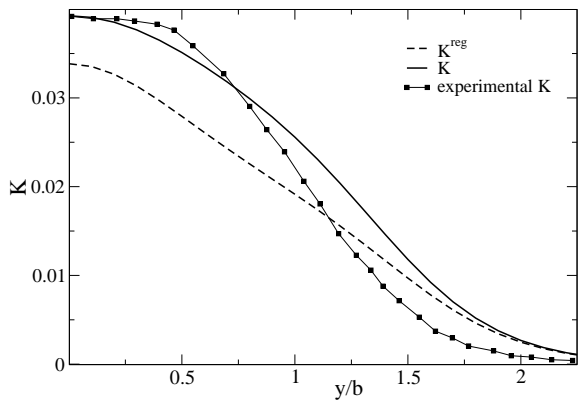}\vskip 1.2cm
 \caption{ Experimental profile~\cite{00GT} of the mean square of the
cross-stream component of the fluctuating velocity (dots connected by line). In a continuous line
we present the profile predicted by the random model. The dashed line is the profile of the
regular contribution. The difference at $a$ is the source of spreading of the jet, cf. Eq. (\ref{angle}).}
\label{RMS}
\end{figure}

To find  the spreading angle $\alpha$ we  should consider the normalized  velocity fluctuations
in the transverse direction $y$,
 \begin{subequations} \label{K} \begin{eqnarray}
  K (x,y) &\= & \< \overline{ U_{y }-V_{y })^2} \>   = \< \overline{U_ y^2} \> -V_y^2\,, \\
  K^\ddag(\zeta) &\=&    K(x,y)\big / V_x(x,0)^2  \ .
\end{eqnarray}\end{subequations}
One simple way to introduce randomness in the model is to account for
the fluctuations of the core size of the vortices, $c_n=c\,
(1+\delta_n)$ with uncorrelated statistical fluctuations:
$ \overline{\delta_n}=0\,, \quad  \overline{\delta_n \delta_{n'}}
\equiv \delta^2\Delta_{n,n'}$.
 In the limit $\alpha\to 0$ we replace the parameter $c$ in $\C R^2_{n}$ 
 by $c_n$ and (for small fluctuations, $\delta^2\ll 1$ ) expand \Eqs{actU}
 with respect of $\delta_n$.  Averaging the result with respect to the
 fluctuations and over longitudinal position in the jet we get:
\begin{subequations}\label{U2}
\begin{eqnarray}\label{U2B}
\<\overline{U_y^2}\>&=&\(\frac{\gamma}{2\, \pi \, b} \)^2\frac 1 {2d} \int\limits _{-d}^d
d\xi \sum_{n,n'=-\infty}^\infty \C X_n \C X_{n'} E_{n,n'}\,, \\ \label{U2C}
E_{n,n'}&\=& E_{n,n'}\sp{reg}+E_{n,n'}\sp{ran}\,, \quad  E_{n,n'}\sp{reg}\= \frac 1{\C R^2_n \C R^2_{n'}}\,, \\ \nonumber E_{n,n'}\sp{ran}&\=&2\, \delta^2\Big\{\frac {2 c^6 \Delta_{n, n'}}{\C R_n^{12}}  + \frac{3\, c^2\big [c^4-(\C X_n^2+\C Y_n^2)^2\big]}{\C R_n^{10} \C R^2_{n'}}\Big\}\ .
\end{eqnarray}
\end{subequations}
\noindent In accordance with these results we present $K$ as a sum
of the regular part $K\sp{reg}$, that originates from $x$-periodic velocity fluctuations, and the random part $K\sp{ran}$, that originates in our model from statistical
fluctuations of the core radius: $K=K\sp{reg}+K\sp{ran}$.  Here
$K\sp{reg}\equiv \<\overline{U_y^2}\>\sp{reg}-V_y^2$ and
$K\sp{ran}\equiv \<\overline{U_y^2}\>\sp{ran}$, where
$\<\overline{U_y^2}\>\sp{reg} $ and $\<\overline{U_y^2}\>\sp{ran} $
are related in \Eqs{U2} with $E_{n,n'}\sp{reg}$ and
$E_{n,n'}\sp{ran}$.  Experimental information about $K$ is provided in
Fig. \ref{RMS}.  We see from the data that on the axis of the jet
$K^\ddag\approx 0.04$ and at the centers of the vortices $K^\ddag(\zeta=a)\approx 0.03$. We use this data to fix the parameters of the model with the results $d=1.45$ and
$\delta^2= 0.03$.  With these parameters the regular contribution $K\sp{reg}(a)\approx 0.024$ while the random contribution is considerably smaller, $K\sp{ran}(a)\approx 0.006$.

 In this way we have separated the regular contribution to $K$ that
 does not effect  the spreading angle, from the random one,
 $K\sp{ran}$, that leads to the jet spreading by turbulent diffusion with
 a characteristic velocity $\sqrt{K\sp{ran}}$. We thus estimate
 \begin{eqnarray}
 \alpha\sb{mod}\simeq \frac{\sqrt{K\sp{ran}(a)} } {V_x(a)}= \frac{\sqrt{{K_y^\ddag}\sp{ran}(a)}}{V^\ddag_x(a)}
 \approx 0.13  \ . \label{angle}
 \end{eqnarray}
 This estimate of the tangent of the angle of the jet is
 slightly higher than the experimental value (about 0.1). We note
 however that we made no effort to model quantitatively the random
 structure of the vortices. For example, looking at
 Fig. \ref{structure2} we note that the vortices are elliptic rather
 than round, this will definitely contribute to lowering the random
 part of $K^\ddag$. One could input this knowledge, with a price of
 additional fit parameters. We submit to the reader that this is not
 the goal of the present calculation, we are not interested here in a
 quantitative model of a jet spreading. We wanted to understand what
 is the physical reason for this phenomenon and why the jet angle is
 small. We believe that the model worked out above is sufficient; our
 answer is that any deterministic model will necessarily close the jet
 angle, allowing only a self similar parallel jet. The reason for the
 opening angle is randomness, which is the flip side of turbulence
 which cannot be avoided even when 75\% of the energy is in the
 coherent structure. The randomness contributes partially to the value
 $K^\ddag$, which is rather small per se.  Therefore
 ${K^\ddag}\sp{ran}$ (without which the opening angle will tend to
 zero) is even smaller and thus resulting spreading angle~\eq{angle}
 is indeed small.\\ 

 \noindent {\bf Summary}.  In this Letter we offered a simple model of
 a plane jet with the aim of understanding the fundamental existence
 of a statistically self similar structure with a small opening
 angle. The deterministic part of the model excellently reproduces the
 experimental profiles of the mean velocity without using any
 adjustable parameter. For an opening angle we must introduce
 randomness. In the context of the present simple model we ascribed
 the randomness to the core size, and demonstrated that this is
 sufficient to provide an opening angle of the correct order of
 magnitude. Needless to say we do not pretend that this model describes correctly
 the full randomness which is due to turbulence, but only underlines the crucial
 role of the random components in opening up the jet. One can provide a more precise model with better agreement with the measured angle and the profiles of second order quantities,
 but this calls for additional adjustable parameters. Such a detailed
 model is not the aim of this Letter which concentrated on
 understanding the basic physics of the phenomenon.\\ 

 \noindent {\bf Acknowledgement}. IP thanks Roddam Narasimha for presenting this riddle,
 and acknowledges partial support by the US-Israel BSF. Additional support 
was given by the Transnational Access Programme at RISC-Linz, funded
by European Commission Framework 6 Programme for Integrated
Infrastructures Initiatives under the project SCIEnce (Contract No.
026133).


\begin{thebibliography}{99}
\bibitem{footnote} By definition the half jet width $b(x)$ is the
cross-stream position at which the stream-wise mean velocity is half
of its center-line value.
\bibitem{note1} Equation for $V_x^\ddag(\zeta)$ can be obtained from
 \Eq{U} similarly to \Eq{PVS}. For the profile \eq{Ux-fin} and a given $\alpha $ the values of $a$ and $c$
were found by the procedure described in the text. It turns out that for small
$\alpha$  ($\lesssim 0.15$) the values of $a$ and $c$ practically coincide with
the corresponding values for $\alpha=0$.
 In the limit $\alpha\to 0$
 the main contribution to the integrals $J^{\pm}(\zeta)$ is due to the
 region $\chi\approx 1$. Introducing in \Eq{Ux-fin} a dummy variable
 $\xi=(\chi-1)/\alpha$ one sees that $\displaystyle \lim_{\alpha\to
 0}J^{\pm} (\zeta)=I^{\pm} (\zeta)$ and profile $V_x^\ddag (\zeta)\to \C
 V_x^\ddag(\zeta)$.
\bibitem{note2} Using the Prandtl closure $W=\nu_{_{\rm T}} dU_x/dy$
in the balance equation for the mechanical moment and assuming that
the turbulent viscosity $\nu_{_{\rm T}}$ is constant in the jet, one
predicts a universal, $\nu_{_{\rm T}}$-independent profile, see
e.g. Eq.~(5.187) in \cite{Pope}
\end{thebibliography}
\end{document}